\documentclass[aps,final,notitlepage,oneside,nobibnotes,nofootinbib,superscriptaddress,noshowpacs,centertags,showpacs]{revtex4-1}

\usepackage[english]{babel}
\usepackage{graphicx}
\usepackage{latexsym}
\usepackage{amssymb}
\usepackage{amsmath}

\begin{document}

\title{Quantum correction to black hole accretion}

\author{V.\;I. Dokuchaev}\thanks{e-mail: dokuchaev@inr.ac.ru}
\affiliation{Institute for Nuclear Research of the Russian Academy of Sciences
60th October Anniversary Prospect 7a, 117312 Moscow, Russia}
\affiliation{Moscow Institute of Physics and Technology, 9 Institutskiy per., Dolgoprudny, 141701  Moscow Region Russia}

\date{\today}

\begin{abstract}
We describe quantum correction to the accreting hot plasma onto black holes. This quantum correction is related with the Hawking radiation, which heats the accreting plasma. The hot accreting gas is heated additionally by the quantum Hawking radiation. It is demonstrated that Hawking radiation prevails over the Compton scattering of hot electrons in the accreting flow onto the small enough evaporating black holes with masses $M<M_q\simeq 4.61\cdot10^{29}$~grams. In result, the evaporating black holes with masses $M<M_q$ reverse the inflowing plasma into outflowing one and stop the black hole accretion at all. The black holes with masses $M<M_q$ made contribute to the enigmatic dark matter at the galactic disks, galactic halos and even in the intergalactic space, if these black holes are primordial in origin.
\end{abstract}


\maketitle


\section{Introduction}

The  hot plasma accreting onto black hole is heated additionally and inevitably by the outgoing particles of the quasi-classical quantum Hawking radiation. It the following we allege that Hawking radiation would prevail over the Compton scattering of hot electrons in the accreting flow onto the small enough evaporating black holes. 

It would be demonstrated that small enough black holes may by a part of dark matter at the galactic disks, galactic halos and even in the intergalactic space, if these black holes are primordial in origin.

We provide all necessary analytical and numerical calculations of the test particle motion, both massive and massless, in the Kerr--Newman metric.

\section{Kerr--Newman metric}

The line element of the classical Kerr--Newman metric~\cite{Kerr,Newman1,Newman2, Carter68, mtw,Chandra}, describing the rotating and electrically charged black hole, is  
\begin{equation}
	ds^2=-e^{2\nu}dt^2+e^{2\psi}(d\phi-\omega dt)^2 +e^{2\mu_1}dr^2+e^{2\mu_2}d\theta^2,
	\label{metric}
\end{equation}
where
\begin{eqnarray}
	e^{2\nu}&=&\frac{\Sigma\Delta}{A}, \quad e^{2\psi}=\frac{A\sin^2\theta}{\Sigma}, 
	\quad e^{2\mu_1}=\frac{\Sigma}{\Delta}, \quad e^{2\mu_2}=\Sigma, \quad
	\omega=\frac{2Mar}{A}, \label{omega} \\
	\Delta &= & r^2-2Mr+a^2+q^2, \quad \Sigma=r^2+a^2\cos^2\theta, \quad A=(r^2+a^2)^2-a^2\Delta\sin^2\theta. \label{A}
\end{eqnarray}
In these equations $M$ --- black hole mass, $a=J/M$ \ --- \  black hole specific angular momentum (spin), $q$ \ --- \  black hole electric charge, $\omega$ \ --- \ frame-dragging angular velocity. We often use units with the gravitational constant $G=1$ and the velocity of light $c=1$. For simplification of formulas in the following, we often use the dimensional values for space distances $r\Rightarrow r/M$, for time intervals $t\Rightarrow t/M$ and etc. In other words, we will measure the radial distances in units $GM/c^2$ and time intervals in units $GM/c^3$. We also will use the dimensionless value for black hole spin $a=J/M^2\leq1$, by supposing that $0\leq a\leq1$. The black hole event horizon radius $r_+$ and the Cauchy radius $r_-$ are the roots of quadratic equation $\Delta=0$:
\begin{equation}
	r_{\pm}=1\pm\sqrt{1-a^2-q^2}.
	\label{r+-}
\end{equation}
The specific feature of the rotating Kerr--Newman black hole is the solid or rigid rotation (i. e., independent of the polar angle $\theta$) of the black hole event horizon with the angular velocity 
\begin{equation}
	\omega_+=\frac{2Mar_+}{(r_+^2+a^2)^2}
	\label{omega+} 
\end{equation}
For test particles in the Kerr--Newman metric there are four integrals of motion: $\mu$ ---  particle mass, $E$ --- particle total energy, $L$ --- particle azimuth angular momentum and $Q$ --- Carter constant, related with non-equatorial motion \cite{Carter68}. The radial potential $R(r)$ with these integrals of motion defines the radial motion of particles:
\begin{equation}
	R(r) = P^2-\Delta[\mu^2r^2+(L-aE)^2+Q],
	\label{Rr} 
\end{equation}
where $P=E(r^2+a^2)-a L$. At the same time, the polar potential $\Theta(\theta)$ defines the motion of particles in the polar direction:
\begin{equation}
	\Theta(\theta) = Q-\cos^2\theta[a^2(\mu^2-E^2)+L^2\sin^{-2}\theta].
	\label{Vtheta} 
\end{equation}

All trajectories of massive test particles ($\mu\neq0$) depend on three parameters (constants of motion or orbital parameters): $\gamma=E/\mu$, $\lambda=L/E$ and $q=\sqrt{Q}/E$. Respectively, the corresponding trajectories of massless particles ($\mu\neq0$) depends only on two parameters: $\lambda=L/E$ and $q=\sqrt{Q}/E$. 

For a static distant telescope (observer), placed at the radius $r_0\gg r_{\rm h}$ (e.\,g., at the asymptotically flat part of the space-time), at the polar angle $\theta_0$ and at the  azimuth angle $\phi_0$, the horizontal impact parameter $\alpha$ and vertical impact parameter $\beta$ must be used on the celestial sphere (see details in~\cite{Bardeen73,CunnBardeen72,CunnBardeen73}): 
\begin{equation}
	\alpha =-\frac{\lambda}{\sin\theta_0}, \quad
	\beta = \pm\sqrt{\Theta(\theta_0)},
	\label{alpha} 
\end{equation}
where the effective polar potential  $\Theta(\theta)$ is from equation (\ref{Vtheta}). 

\begin{figure}
	\centering
	\includegraphics[width=0.46\textwidth]{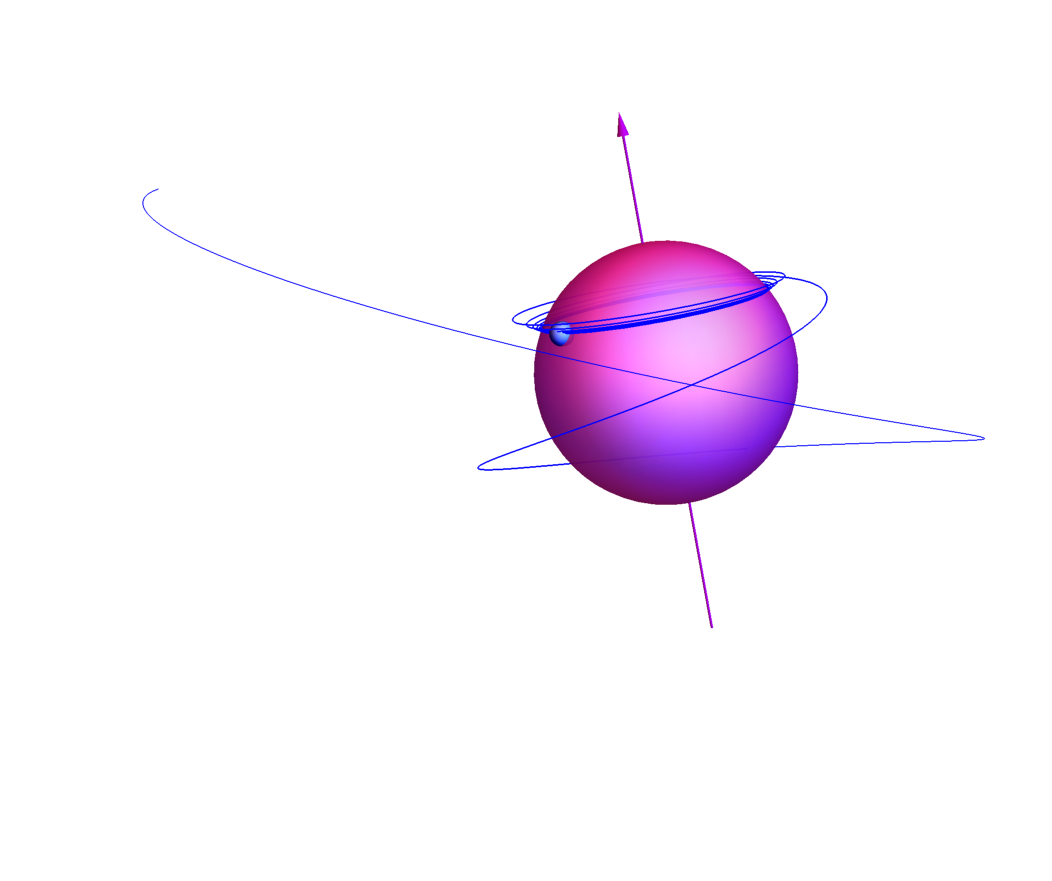}
	\hfill
	\includegraphics[width=0.48\textwidth]{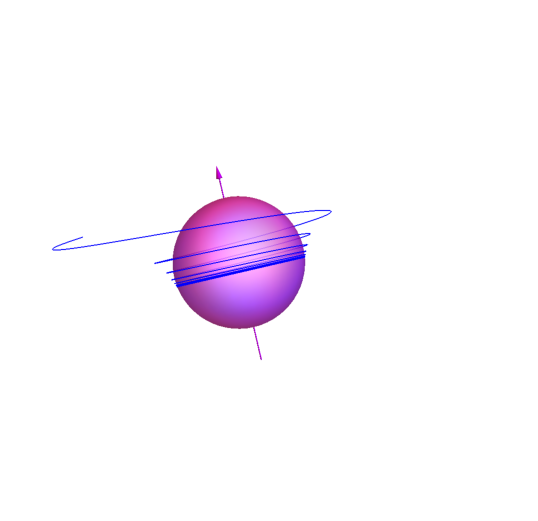}
	\caption{Massive ($\mu\neq0$) test particle (or planet) plunging into a fast-rotating Kerr black hole with spin $a=0.998$. \textbf{{Left panel}}: Particle orbital parameter $Q=1$, $\gamma=0.85$ and $\lambda=1.7$.  \textbf{{Right panel}}:  Particle orbital parameter $Q=0.4$, $\gamma=0.85$ and $\lambda=1.7$. Near the black hole event horizon all particles are winding around black hole horizon globe with a constant angular velocity $\omega_+$, which does not depend on the polar angle $\theta$. (See, e.\,g., \cite{dokuch14,dokuch19,doknaz19b} for details of analytical and numerical calculations of test particles trajectories.)}
	\label{fig1}
\end{figure}

\begin{figure}
	\centering
	\includegraphics[width=0.38\textwidth]{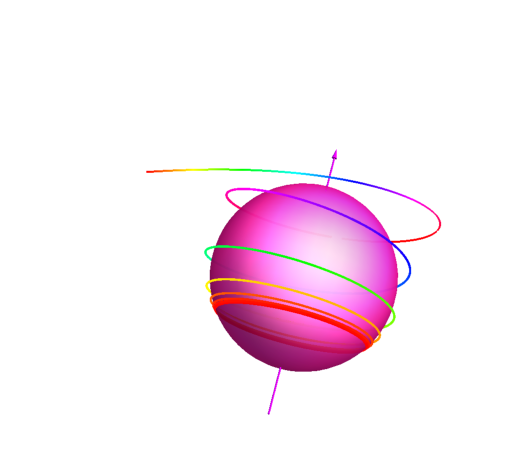}
	\hfill
	\includegraphics[width=0.58\textwidth]{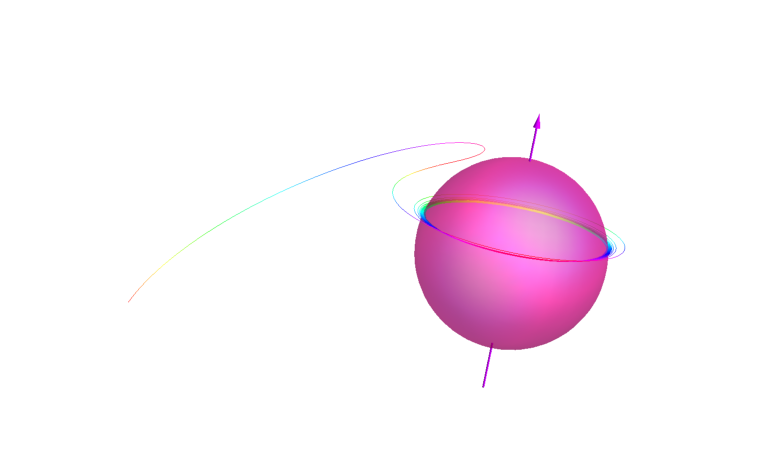}
	\caption{Photons plunging into a fast-rotating Kerr black hole. \textbf{{Left panel}}: black hole spin $a=0.9982$, photon orbital parameters $Q=2$ and $\lambda$=2. \textbf{{Right panel}}:  black hole spin $a=1$, photon orbital parameters $Q=4$ and $\lambda$=-6.5. Near the black hole event horizon photons are winding around black hole horizon globe with a constant angular velocity $\omega_+$, which does not depend on the polar angle $\theta$ according to equation (\ref{omega+}).}
	\label{fig2}
\end{figure}

At Fig.~\ref{fig1} are shown two trajectories of test massive particles ($\mu\neq0$) plunging into a fast-rotating black hole. Note, that by approaching the black hole horizon, these trajectories are winding up around the event horizon globe with the constant angular velocity $\omega_+$ from equation (\ref{omega+}). This winding is the general properties of all trajectories plunging into rotating black hole as it is illustrated at Fig.~\ref{fig2} for the corresponding trajectories of massless particles ($\mu=0$) like photons. 

\begin{figure}
	\centering
	\includegraphics[width=0.4\textwidth]{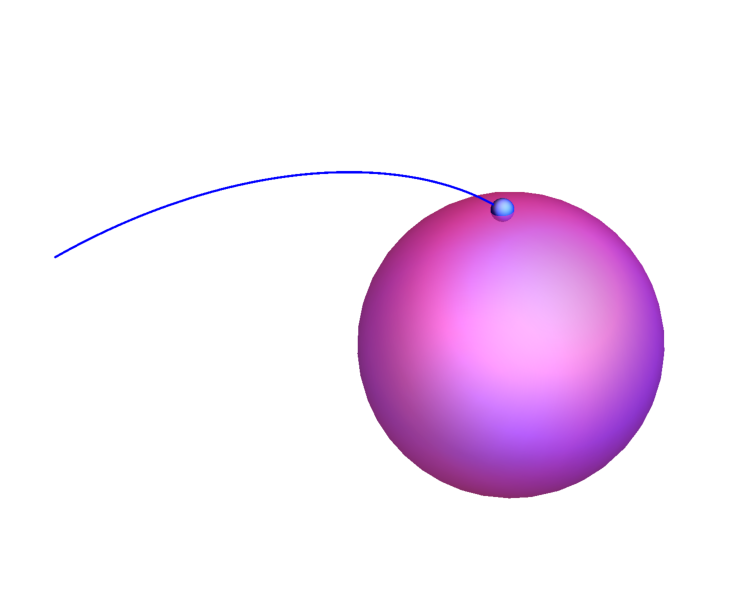}
	\caption{Massive ($\mu\neq0$) test particle (or planet) with orbital parameters $Q=3$, $\gamma=0.94$ and $\lambda=1.3$, plunging into a spherically symmetric non-rotating Schwarzschild black hole with spin $a=0$.}
	\label{fig3}      
\end{figure}

At last, Fig.~\ref{fig3} shows the trajectory of massive particle ($\mu=0$) plunging into the Schwarzschild black hole, which is both spherically symmetric and nonrotating ($a=0$). Note that we calculated analytically and numerically all trajectories of massive and massless test particles, presented at Figs~\ref{fig1}, \ref{fig2} and \ref{fig3}, by using Carter equations of motion \cite{Carter68} in the Kerr--Newman metric (see for details, e.\,g., \cite{dokuch14,dokuch19,doknaz19b}). 

From astrophysical point of view (see, e. g., \cite{Ayzenberg23}) the most interesting are the cases of fast-rotating black holes with spin values close to the maximum value, $a_{\rm max}=1$. 

\section{The quantum correction to the black hole accretion}
\label{classical}

Stephen Hawking recovers the thermal radiation of black holes in the quasi-classical approximation \cite{Hawking74,Hawking75,BCH} with the temperature 
\begin{equation}
	T_+={\frac{\kappa_+}{2\pi}},
	\label{T+}
\end{equation}
where the so-called "{\it surface tension}"{} of the black hole horizon $\kappa_+$ in the Kerr--Newman metric \cite{Kerr,Newman1,Newman2, Carter68, mtw,Chandra} is  
\begin{equation}
	k_+=(r_+ -r_-)/(2(r_+^2 +a^2))= \frac{\sqrt{M^2-q^2-a^2}}{2M^2-q^2+2M(M^2-q^2- a^2)^{1/2}}.
	\label{k+}
\end{equation}
In the Schwarzschild case ($a=0$ and $q=0$)
\begin{equation}
	T_+=\frac{\hbar c^3}{8\pi GM k_B},
	\label{T+S}
\end{equation}
where $k_B$ is the Boltzmann constant.  

The temperature of Hawking radiation tends to zero, while black spin $a$ tends to the extreme (or maximal) value $a_{\rm max}=1$. Meanwhile, the fast-rotating Kerr black holes spontaneously emit waves (in particular, the scalar waves) with the intensity close to the Hawking radiation for the nonrotating black hole \cite{Starobinskii}. For this reason we use the Schwarzschild case in our estimation of the quantum correction to the black hole accretion.

The useful approximation to the spherical black hole accretion is the so called "{\it Bondi accretion}" in which
the accretion onto spherically symmetric Schwarzschild black hole is modeled by the accretion of adiabatic gas with the  adiabatic index $\gamma=3/2$, local gas pressure and gas density, respectively, $p_\infty$ and $\rho_\infty$ far from the black hole, and local gas sound speed far from the black hole $c_s=\gamma p_\infty/\rho_\infty$. 

Spherically accreting black hole influences the gas motion inside the  so-called "{\it influence}"{} radius, which is called also a Bondi radius $r_B$ \cite{Bondi}, defined as
\begin{equation}
	r_B\equiv\frac{GM}{c_s^2}\rho_\infty.
\end{equation}
The corresponding modeled Bondi accretion rate $\dot M_B$ is approximately
\begin{equation}
	\dot M_B\approx2\pi\frac{(GM)^2}{c^3}\rho_\infty.
	\label{BondiRate}
\end{equation}

The hot acreeting plasma is heated additionally by the outgoing photons of the Hawking radiation. It is clear that Hawking radiation would prevail over the Compton scattering of hot electrons in the accreting flow onto black hole if $r_+<h/(m_e c)$ or
\begin{equation}
	M<M_q\simeq\frac{h}{mc}\frac{c^2}{G}=\frac{M_{Pl}^2}{m} \simeq4.61\cdot10^{29} \: \mbox{gram},
	\label{quantum}
\end{equation}
where $M_{Pl}\simeq\sqrt{hc/G}=2.176\cdot10^{-5}$~gram is the Plank mass and $m_e\simeq2.109\cdot10^{-34}$~gram is the electron mass, respectively. In result, the quantum correction to black hole accretion is crucial for the small enough evaporating black holes.

\section{Discussions and Conclusions}

The hot plasma accreting onto black hole is heated additionally by the outgoing photons of the quantum Hawking radiation, especially due to a winding of accreting particles near the black hole event horizon, as demonstrated  at Figs.~1 and 2. 

The quantum Hawking radiation prevails over the Compton scattering of hot electrons in the accreting flow onto black hole if the mass of evaporating black hole is small enough. 

Namely, the small enough evaporating black holes with masses $M<M_q$ from equation (\ref{quantum}) reverse the inflowing plasma into outflowing one and stop the black hole accretion at all. 

The black holes with masses $M<M_q$ made contribute to the enigmatic dark matter at the galactic disks, galactic halos and even in the intergalactic space, if these black holes are primordial in origin. 


\end{document}